\documentclass[
preprint, showpacs, showkeys
 amsmath,amssymb,
 aps,
prc,
 longbibliography,
 lengthcheck,%
]{revtex4-1}

\usepackage{graphicx}% Include figure files
\usepackage{dcolumn}% Align table columns on decimal point
\usepackage{bm}% bold math
\usepackage{hyperref}% add hypertext capabilities
\begin{document}

%\preprint{APS/123-QED}

\title{Dynamical approach to spectator fragmentation in \\ Au+Au reactions
at 35 MeV/A}

\author{Yogesh K. Vermani}
\author{Rajeev K. Puri}%
\affiliation{Department of Physics, Panjab University, Chandigarh
-160 014, India }

% \collaboration{MUSO Collaboration}%\noaffiliation

%\affiliation{
%\affiliation{
%\affiliation{%

\date{\today}% It is always \today, today,

\begin{abstract}
The characteristics of fragment emission in peripheral
$^{197}$Au+$^{197}$Au collisions 35 MeV/A are studied using the two
clusterization approaches within framework of \emph{quantum
molecular dynamics} model. Our model calculations using
\emph{minimum spanning tree} (MST) algorithm and advanced
clusterization method namely \emph{simulated annealing clusterization algorithm} (SACA)
showed that fragment structure can be realized at an earlier time when spectators contribute
significantly toward the fragment production even at such a low
incident energy. Comparison of model predictions with experimental
data reveals that SACA method can nicely reproduce the fragment
charge yields and mean charge of the heaviest fragment. This
reflects suitability of SACA method over conventional
clusterization techniques to investigate spectator matter
fragmentation in low energy domain.
\end{abstract}

\pacs{25.70.-z 25.70.Mn 24.10.Lx}
\keywords{heavy-ion collisions, multifragmentation, quantum
molecular dynamics (QMD) model, Multics-Miniball array}

\maketitle

%\tableofcontents

\section{\label{intro}Introduction}

\par The study of heavy-ion (HI) reactions at intermediate energies
provides an important platform to probe the highly non-equilibrium
environment produced in the reaction zone as well as properties of
excited fragments evolved from the spectator zone
\cite{ogi,beg,kreut,poch,schut}. The spectator matter
fragmentation at relativistic bombarding energies is also characterized 
by a rise and fall pattern \cite{ogi,beg,kreut,schut}. These experiments performed on ALADiN
set-up mainly focussed on the liquid-gas phase transitions
\cite{kreut,poch,schut,fevre} and universality behavior observed
in the fragment-emission at incident energies $\geq~400$ MeV/A
\cite{beg,schut}. The clusterization approach based upon
minimization of fragments' total energy dubbed as \emph{simulated
annealing clusterization algorithm} (SACA) is reported to explain
this rise and fall trend in the multiplicity of IMFs with
collision geometry quite accurately at relativistic bombarding
energies \cite{euro}. Conventional clusterization approach based
upon spatial correlation among nucleons, however, failed
completely to explain higher IMF yields at large impact parameters
\cite{euro,tsang}.

The reaction dynamics and associated non-equilibrium aspects in
low-energy HI collisions are, however, still poorly understood
phenomena \cite{colon,garcia}. The important question associated
with this domain is whether nuclear system can reach thermal
equilibrium before break-up or not \cite{nayak}. Most of
nucleon-nucleon collisions are Pauli blocked and mean field
governs the reaction dynamics in low energy domain
\cite{moli,breng,aich}. In case of peripheral collisions, we have
projectile-like and target-like remnants and fusion-fission events
dominate the scenario. The fusion events disappear as the incident
energy increases marked by the onset of multifragmentation. In the low energy range (\emph{i.e.} between 40 and 70 MeV/A), 
a significant fraction of the intermediate mass fragments originates from mid-velocity
region \cite{colon}. This type of pre-equilibrium emission has
been conjectured as `extended neck' emission due to the dynamical
fluctuations that increase with the incident energy. Many
experiments have indicated that binary dissipative collisions
(BDC) dominate the scenario in such low energy HI collisions
\cite{peter}. As far as decays from quasiprojectile and
quasitarget are concerned, molecular dynamics approaches coupled
with conventional clusterization algorithm mayn't reproduce
fragment charge yields accurately \cite{hagel}.

In the present paper, we aim to see whether SACA method can
describe spectator matter fragmentation in low energy domain or
not. The confrontation of the theoretical predictions employing
advanced clusterization technique with experimental data can be of
importance to understand the physical scenario behind cluster
production mechanism in low-energy HI collisions. We shall compute
the fragment observables for peripheral Au(35 MeV/A) + Au
collisions employing MST and SACA clusterization subroutines. A
comparison with experimental results recently obtained by
Multics-Miniball group \cite{exp,ago} is also attempted to explore
the applicability of SACA method. Section \ref{model} describes
the main features of the QMD model along with simulated annealing
clusterization algorithm (SACA). Our results are discussed in
section \ref{results} and summarized in section \ref{summary}.

\section{\label{model} The model}
\par The \emph{quantum molecular dynamics} model is \emph{A}-body
transport theory that incorporates the quantum features of Pauli
blocking and stochastic nucleon-nucleon (\emph{n-n}) scattering
\cite{aich,sanj}. Each nucleon in the colliding system is
represented by a gaussian wave packet as \cite{aich}:
\begin{eqnarray}
{\psi}_i({\bf r},{\bf p}_i(t),{\bf r}_i(t))=\frac{1}{(2\pi
L)^{3/4}} exp \left[ \frac{i}{\hbar} {\bf p}_i(t)\cdot{\bf
r}\right. \nonumber \\
 \left. -\frac{({\bf r}-{\bf r}_i(t))^2}{4L} \right]. \label{s1}
\end{eqnarray}
Mean position ${\bf r}_{i}(t)$ and mean momentum ${\bf p}_{i}(t)$
are the two time dependent parameters. The gaussian width has
fixed value $\sqrt{L}$=1.08 $fm$. The centers of these Gaussian
wave packets propagate in coordinate (${\cal R}_3$) and momentum
(${\cal P}_3$) space according to the classical equations of
motion: \\
\begin{equation}
\dot{{\bf p}}_i=- \frac{\partial \langle {\cal H}
\rangle}{\partial {\bf r}_i}; ~\ \dot{{\bf r}}_i=\frac{\partial
\langle {\cal H} \rangle}{\partial {\bf p}_i}.\label{euler} \\
\end{equation}
The Hamiltonian ${\cal H}$ appearing in Eq. \ref{euler} has
contribution from the local Skyrme-type, Yukawa and effective
Coulomb interactions \cite{aich}. Since QMD model gives just the
phase space of nucleons, one needs secondary algorithm to
clusterize the phase space. The standard clusterization approach
namely \emph{minimum spanning tree} (MST) procedure assumes that
the correlating nucleons belong to the same fragment if their
inter-nucleon distance $\mid{\bf r}_{i}-{\bf r}_{j}\mid$ is
smaller than 4 $fm$. This approach is successful when the system
is dilute and clusters are well separated in ${\cal R}_3$ space.

\subsection{\label{esaca} The SACA formalism}

\par This sophisticated clusterization approach allows early identification of
fragments before these are well separated in coordinate space. The
SACA method works on the principle of energy minimization of
fragmenting system. The pre-clusters obtained with the MST method
are subjected to a binding energy check \cite{saca,jai}:
\begin{eqnarray}
\zeta_{i}=\frac{1}{N_{f}}\sum_{\alpha=1}^{N_{f}}
\left[\sqrt{\left(\textbf{p}_{\alpha}-\textbf{P}_{N_{f}^{c.m.}}
\right)^{2}+m_{\alpha}^{2}}-m_{\alpha} \right. \nonumber \\
+\left.\frac{1}{2}\sum_{\beta \neq \alpha}^{N_{f}}V_{\alpha \beta}
\left(\textbf{r}_{\alpha},\textbf{r}_{\beta}\right)\right]<
-E_{bind}, \label{be}
\end{eqnarray}
with $E_{bind}$ = 4.0 MeV if $N_{f}\geq3$ and $E_{bind} = 0$
otherwise. In Eq. (\ref{be}), $N_{f}$ is the number of nucleons in
a fragment and $\textbf{P}_{N_{f}}^{c.m.}$ is the center-of-mass
momentum of the fragment. The requirement of a minimum binding
energy excludes the loosely bound fragments which will decay at
later stage. To look for the most bound configuration (MBC), we
start from a random configuration which is chosen by dividing
whole system into few fragments. The energy of each cluster is
calculated by summing over all the nucleons present in that
cluster using Eq. (\ref{be}).

Let the total energy of a configuration k be $E_{k}(=
{\sum_{i}}N_{f}\zeta_{i})$, where $N_{f}$ is the number of
nucleons in a fragment and $\zeta_{i}$ is the energy per nucleon
of that fragment. Suppose a new configuration $k^{'}$ (which is
obtained by (a) transferring a nucleon from randomly chosen
fragment to another fragment or by (b) setting a nucleon free, or
by (c) absorbing a free nucleon into a fragment) has a total
energy $E_{{k}^{'}}$. If the difference between the old and new
configuration $\Delta E (= E_{{k}^{'}}-E_{k})$ is negative, the
new configuration is always accepted. If not, the new
configuration $k^{'}$ may nevertheless be accepted with a
probability of $exp (-\Delta E/\upsilon)$, where $\upsilon$ is
called the control parameter. This procedure is known as
Metropolis algorithm. The control parameter is decreased in small
steps. This algorithm will yield eventually the most bound
configuration (MBC).

Since this combination of a Metropolis algorithm with slowly
decreasing control parameter $\upsilon$ is known as
\emph{simulated annealing}, so our approach is dubbed as
\emph{simulated annealing clusterization algorithm} (SACA). For
the present calculations we have employed improvised version of
SACA method in which fragments are confronted against realistic
binding energies instead of constant binding energy check of -4.0
MeV/nucleon \cite{jpg10}. It may be mentioned that this
modification has nearly no effect on final fragment yields and it
explains the ALADiN multifragmentation data \cite{schut} at
relativistic energies quite nicely.

\section{\label{results} Calculations and comparison with data}

We simulate the reactions of Au (35 MeV/A)+Au at six peripheral
geometries using the MST and SACA methods. A \emph{soft} equation
of state ($\kappa=200$ MeV) along with standard energy dependent
\emph{n-n} cross section was used for the simulation of HI
reactions.
\begin{figure}[!t]
%\begin{center}
\includegraphics[scale=0.50, trim= 20  0  0  1] {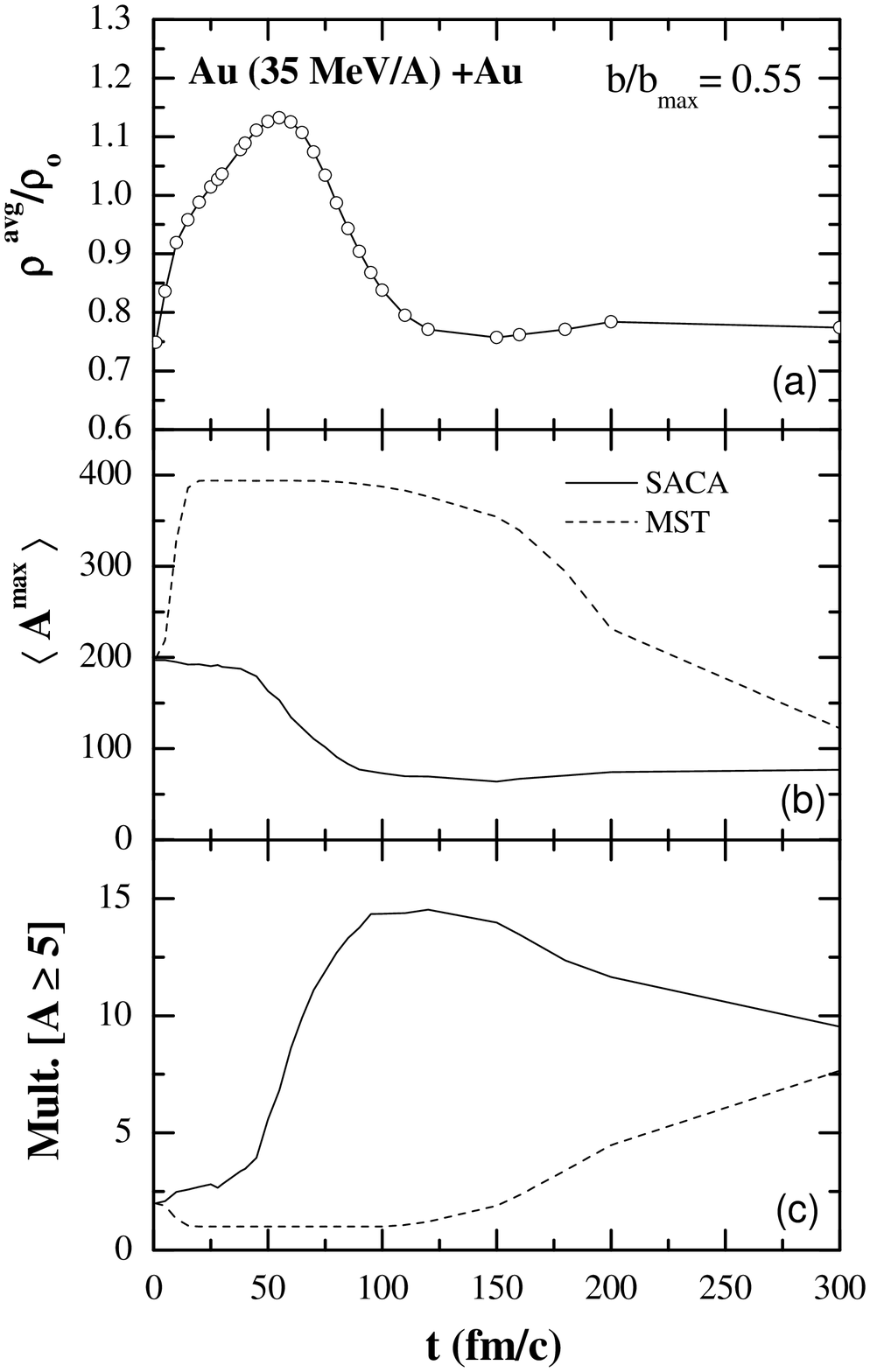}% Here is how to import EPS art
\vskip -0.7cm \caption {The time evolution of (a) mean nucleonic
density $\rho^{avg}/\rho_{o}$; (b) size of the heaviest fragment
$A^{max}$; and (c) the multiplicity of clusters with mass $A \geq
5$. We display the reaction of Au(35 MeV/A)+Au at reduced impact
parameter $b/b_{max}$ =0.55; $b_{max}=1.142[{A_{T}}^{1/3}+
{A_{P}}^{1/3}]$.} \vskip -0.3cm
%\end{center}
\end{figure}

\begin{figure}
%\begin{center}
\vskip 0.8cm
\includegraphics[scale=1.0]{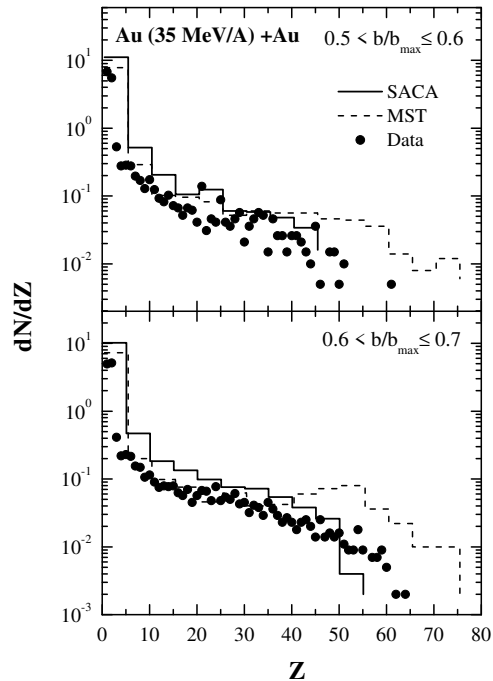}% Here is how to import EPS art
\vskip -0.6cm
\caption {The charge dispersion of the nuclear
fragments in peripheral Au (35 MeV/A)+ Au collisions in the impact
parameter interval $0.5 <b/b_{max}\leq0.6$ (top) and $0.6 <b/b_{max}\leq0.7$ (bottom).
The model calculations using the MST (dotted curve, 300 fm/c) and SACA (solid curve, 100
fm/c) approaches are compared with the experimental data (filled
circles) \cite{exp}.} \label{charge}
%\vskip -0.2cm
%\end{center}
\end{figure}
Figure 1 displays the average density reached in the reaction
along with the evolution of the heaviest fragment $A^{max}$ and
multiplicity of clusters with mass $A\geq5$ followed till 300
fm/c. One can see that as the nucleon density saturates after
violent phase, SACA method is able to identify asymptotic size of
the heaviest fragment $A^{max}$ around 100 fm/c. The QMD+MST
approach, on other hand, fails to detect the heaviest fragment
$A^{max}$ even at 300 fm/c. The heavier $A^{max}$ detected with
MST method at 100 fm/c implies that it still assumes smaller
fragments as constituents of $A^{max}$, being very closely spaced.
We shall show later on in 3-D space that this $A^{max}$ in actual
consists of bunch of smaller clusters. The higher multiplicity of
fragments with mass A$\geq$5 obtained using SACA method (see
figure 1c) clearly shows that $A^{max}$ detected in the MST
approach is actually a bunch of smaller fragments. As shown in
Ref. \cite{jai}, different clusterization approaches converge to
same configuration at 1000 fm/c. It is, however improper to follow
the reaction till 1000 fm/c with semi-classical theory, when
nuclei are found to emit fraction of nucleons even after 100 fm/c.

In figure \ref{charge}, we explore the sensitivity of fragment charge distribution
towards clusterization algorithms. We display here the calculations using MST and SACA
approaches for unfiltered events along with experimental charge yields
(shown as solid circles) obtained from the decay of
the \emph{quasiprojectile} (QP) \cite{exp}. Top and bottom panels show the
comparison of model predictions with experimental data in the impact parameter intervals
$0.5<b/b_{max}\leq 0.6$ and $0.6<b/b_{max}\leq 0.7$, respectively.
The MST method clearly predicts larger production probability for heavier charges.
This method fails to break-up the spectator matter into smaller fragments,
therefore leading to overestimated charge of heavier fragments even at 300 fm/c.
As a result, the tail of charge spectrum shifts towards the higher Z values with
MST method. This also highlights the discrepancy in MST procedure to
describe the spectator fragmentation using semi-classical
transport method. The origin of the fragments with SACA method is,
however, quite earlier determined when nuclear matter is still
excited and compact.

To derive quantitative information on the phenomenon of stopping, one needs to study
distribution of fragments in velocity space. We display in Fig. 3,
the longitudinal rapidity distributions for light charged
particles LCPs [$2 \leq A \leq 4$] and intermediate mass fragments
IMFs [$5\leq A \leq 65$] emitted in reaction of Au(35 MeV/A)+Au at
$b/b_{max}=0.55$. Interestingly, we find that the MST method predicts the
fragment-emission from mid-rapidity zone only (as indicated by a
peak around $(y/y_{beam})_{cm}=0$). SACA approach, on the other hand,
indicates significant contribution coming from the target-like and
projectile-like remnants apart from the mid-velocity emission. MST
method can't cause cause effective break-up of spectator
components to generate the IMFs close to the target and projectile
rapidities even at peripheral geometries. This reflects non-equilibrium
situation in the fragmenting system with SACA approach.
The emission of the LCPs and IMFs near the projectile and target
velocities reflects essentially the binary character of the collisions,
apart from the midrapidity source emission. The peaks observed at the
target and projectile rapidities is also indicative of dynamical scenario
of multi-fragment emission where system has not enough time to pass through
a state of thermodynamical equilibrium. It is worth
mentioning that Ca+Ca collisions at 35 MeV/A were studied recently
within antisymmetrized molecular dynamics (AMD) model
\cite{furut}. Fragment observables such as transverse kinetic
energy of fragments and radial size of the reaction system are
observed to deviate appreciably from the assumption of equilibrium
ensemble. This study also favored dynamical scenario of
fragmentation.
\begin{figure}[!t]
%\vskip -0.8cm
\includegraphics[scale=0.45,trim=12 0  0  0] {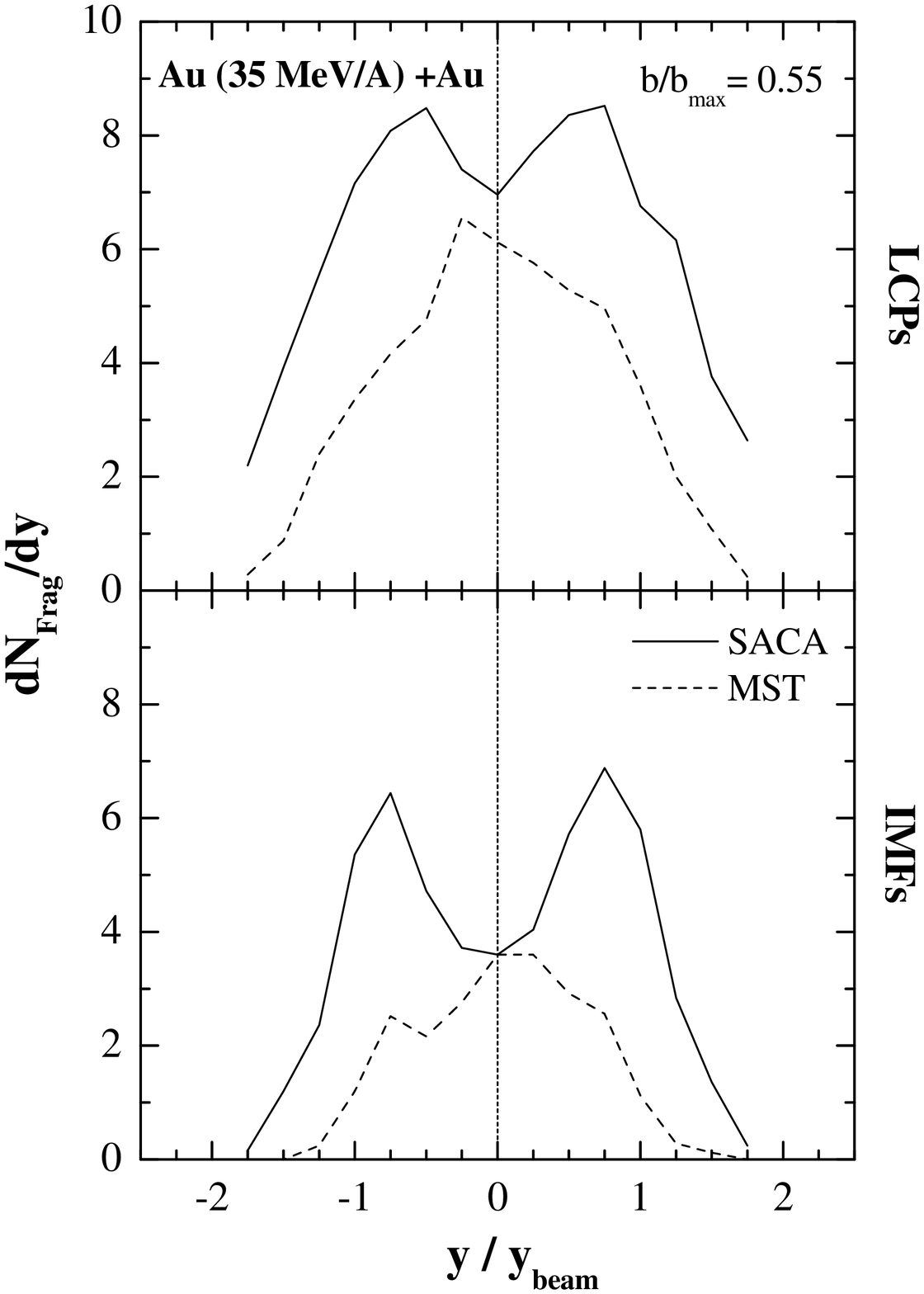}
\vskip -0.8 cm \caption {The longitudinal rapidity distribution
$dN_{Frag}/dy$ of LCPs (top panel) and IMFs (bottom panel)
observed in the reactions of Au (35 MeV/A)+Au at a reduced impact
parameter of $b/b_{max}=0.55$. Our calculations are performed
within MST (dotted curve) and SACA (solid curve) approaches.}
\end{figure}

\begin{figure}
\vskip 1.0cm
\includegraphics[scale=0.49, trim=15 0  0  0]{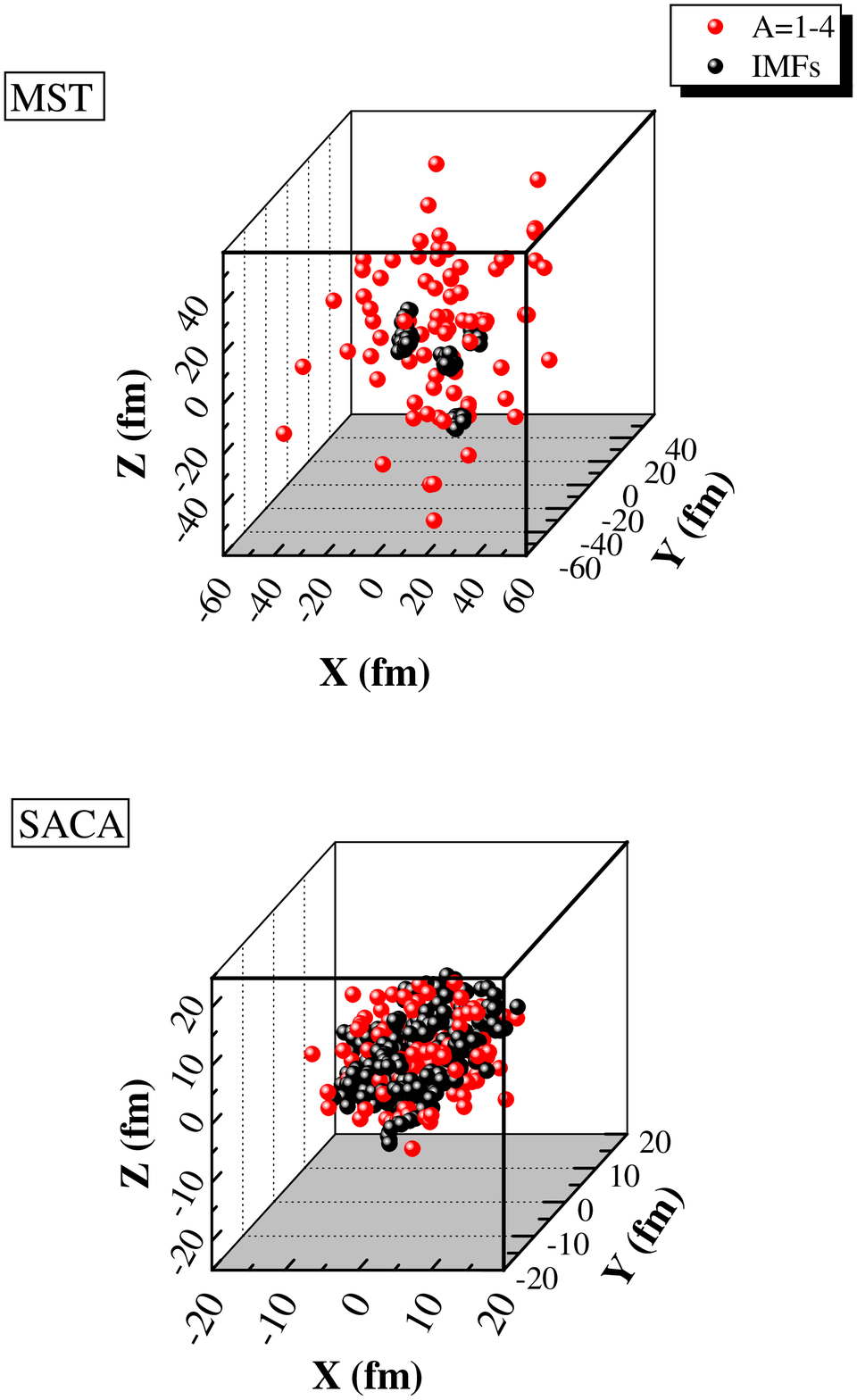}% Here is how to import EPS art
\vskip -0.4cm \caption{\label{size} (Colored online) 3-D snapshots
of a single event of Au (35 MeV/A) + Au at $b/b_{max}=0.55$ using
MST (top) and SACA (bottom) pictures. Black spheres represent the
nucleons distribution in IMFs and red spheres represent the
nucleons distribution in clusters with mass $A=1-4$.}
%\vskip -1.0cm
\end{figure}
Next, we turn to the cluster distribution obtained using MST and
SACA approaches in three-dimensional (3-D) coordinate space.
Figure 4 displays the 3-D snapshots of cluster distribution
obtained in a single event of Au(35 MeV/A) + Au collision at
$b/b_{max}=0.55$. In the MST method, free nucleons and LCPs [$2
\leq A \leq 4$] are abundantly scattered in the whole space (shown
as red spheres) indicating their isotropic emission from the
participant zone. Only a small fraction of intermediate mass
fragment IMFs [$5\leq A \leq 65$] can be seen coming out of the
central overlap region. On the other hand, a significant enhancement in the production of IMFs is
clearly visible for spectator zones using the SACA picture. The contribution towards
the IMFs doesn't seem to come from any specific region. In other
words, the QMD + SACA calculations suggest that IMFs originate
from the `extended neck' region as well as from the spectator
zones. The MST approach fails to break the spectator components
efficiently, thereby under-estimate the IMF yields. These results
highlight the importance of clustering criterion in describing
reaction mechanism in low energy domain of HI reactions.

\begin{figure}[!t]
\vskip -2.5cm
\includegraphics[scale=0.52, trim=20 0  0  0] {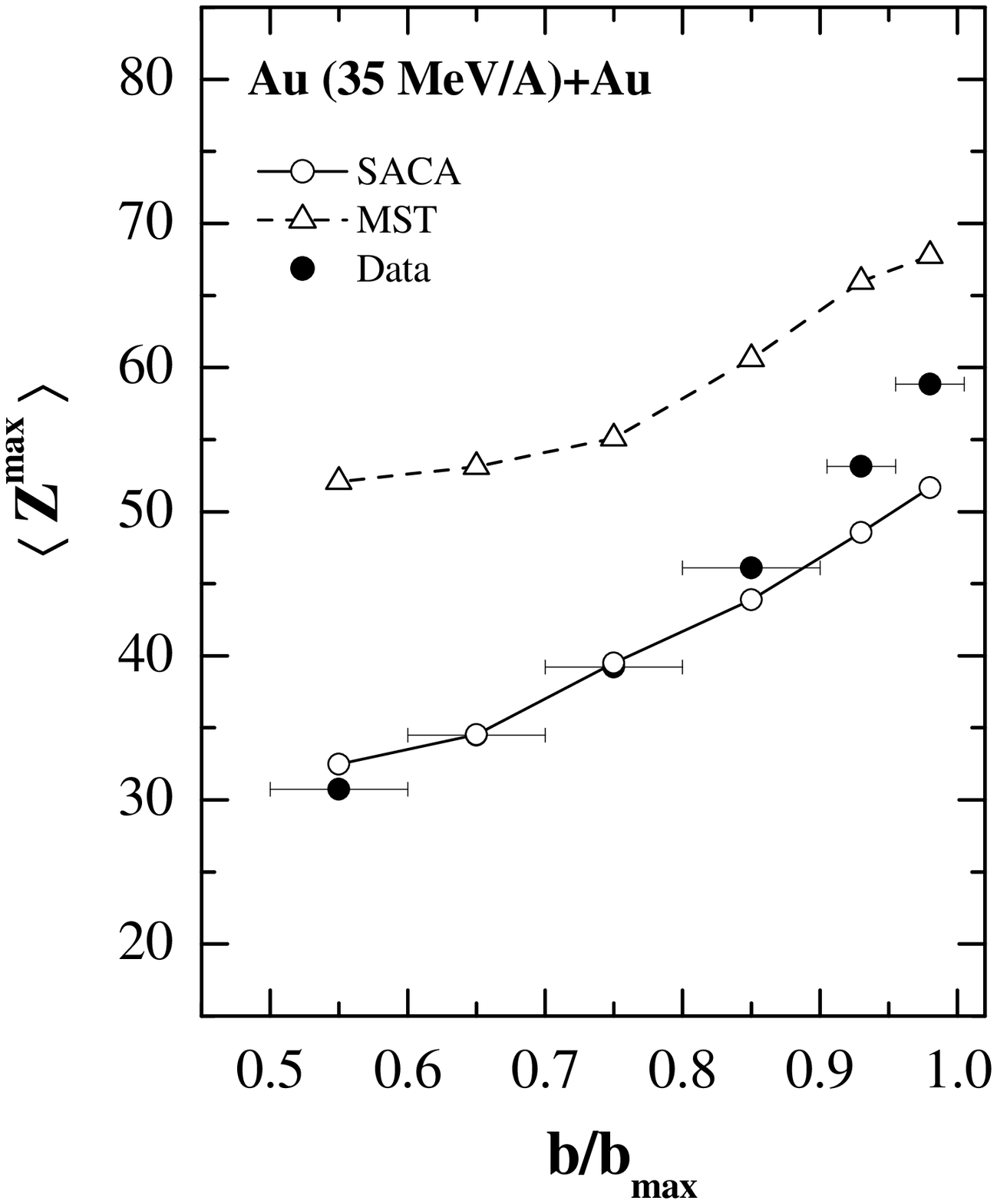}% Here is how to import EPS art
\vskip -3.5cm \caption {The charge of the heaviest fragment
$\langle Z^{max}\rangle$ as a function of the reduced impact
parameter $b/b_{max}$. Our calculations employing the MST (dotted
curve) and SACA (solid curve) approaches are compared with
experimental data (filled circles) \cite{ago}.}
%\vskip -0.4 cm
\end{figure}
Taking the advantage that SACA approach is able to reproduce the
experimental charge yields quite accurately, we further compare
mean charge of the heaviest fragment $\langle Z^{max} \rangle$
obtained using MST and SACA methods with data. Figure 5 shows the
mean charge of heaviest fragment $\langle Z^{max}\rangle$ as a
function of `reduced' impact parameter $b/b_{max}$ for Au (35 MeV/A)+Au reactions.
The model predictions using SACA (at 100 fm/c)
and MST (at 300 fm/c) approaches are displayed along with
experimental data taken with combined Multics-Miniball (MM) array
\cite{ago}. One can see an increasing trend of $Z^{max}$ with
impact parameter as is expected due to increase in size of
spectator zone in the exit channel. Here also, QMD+MST approach
fails to break the spectator matter effectively, and therefore
leads to overpredicted $Z^{max}$ even at 300 fm/c. The SACA
method, on other hand, reproduces the experimental trend quite
accurately at an earlier time. From these finding, we can infer
that the SACA method is on reliable footing to explore the
dynamics of spectator matter fragmentation in low energy heavy-ion
collisions. Further it allows faster recognition of clusters on
reaction time scale.

\section{\label{summary}Summary}
Summarizing, we have performed a comparative study of two
different clusterization models by simulating the peripheral
$^{197}Au$+$^{197}Au$ collisions at 35 MeV/A. Our calculations
within QMD approach coupled with SACA clusterization subroutine
showed that spectator zones contribute significantly towards
fragment production in peripheral HI collisions. Contrary to this,
the MST method could predict fragment-emission from mid-velocity
source only even at 300 fm/c after the initial contact between
projectile and target nuclei. This questions the validity of
employing MST method of clusterization to investigate reaction
dynamics at low incident energies. Our model predictions using the
SACA method reveal that significant contribution towards IMFs
emission is located near the initial target and projectile
rapidities. Our model predictions for the charge distribution and
mean charge of heaviest fragment $\langle Z^{max}\rangle$ using
the SACA method are in nice agreement with experimental data taken
with Multics-Miniball (MM) array \cite{exp,ago}. These findings
highlight the importance of clustering criterion in describing
mechanism behind low-energy spectator fragmentation. The SACA
formalism can, therefore, be treated as general improvement over
conventional clusterization algorithm to recognize the fragment
structure at low and intermediate energies. \\

\begin{acknowledgments}
One of the authors (Y. K. V) is thankful to Drs. M. Bruno and M. D' Agostino at I N F N,
Italy for constructive discussions and interest
shown. The research grant from Department of Science and
Technology, Government of India vide grant no. SR/S2/HEP-28/2006
is gratefully acknowledged.
\end{acknowledgments}

% ****************************************************************8

\end{document}